\documentclass[a4paper,12pt, oneside]{article}
\usepackage{amsthm}
\usepackage{amsmath}
\usepackage{amssymb}
\usepackage{color}
\usepackage{graphicx}
\usepackage{subfigure}
\usepackage{cite}
\usepackage[linktocpage]{hyperref}
\usepackage{setspace}

\usepackage{amssymb}
\usepackage{graphicx}
\usepackage{epsfig}
\usepackage{bm}
\usepackage{mathrsfs}
\usepackage{amsfonts}
\usepackage{epstopdf}
\usepackage{bm}
\usepackage{amsthm}
\usepackage{multirow}
\usepackage{subfigure}
\usepackage{amsmath}
\usepackage{textcomp}
\usepackage{fancyhdr}
\usepackage{mathtools}

\usepackage{indentfirst} 
\usepackage{geometry}
\usepackage{xcolor} 
\usepackage{hyperref}
\hypersetup{
	colorlinks = true,    
	linkcolor  = blue,    
	citecolor  = blue,      
	filecolor  = blue,    
	urlcolor   = blue     
}

\begin{document}
\let\endtitlepage\relax
	
\begin{titlepage}
\begin{center}
			
			\newcommand\blfootnote[1]{%
			\begingroup
			\renewcommand\thefootnote{*}\footnote{#1}%
			\addtocounter{footnote}{-1}%
			\endgroup
}

\vspace*{-1.0cm}
\renewcommand{\baselinestretch}{1.0}  %Line spacing
\setstretch{1.4}
\Large{\textbf{Conserved charges and the first law of black holes for field dependent symmetry generators in the covariant phase space formalism}}

\renewcommand{\baselinestretch}{1.0}  %Line spacing
\setstretch{1.2}
			
\vspace{8mm}
\centerline{\large{Hai-Feng Ding\blfootnote{Email:~haifeng1116@qq.com} }}
\vspace{3mm}
\normalsize
\textit{Institute of Geometric Arts, Chuxiong Normal College, Chuxiong 675000, China}
			
\renewcommand{\baselinestretch}{1.0}  %Line spacing
\setstretch{1}
			
%--------------------------------------------------------------------------------------------------------------------------------------
			
\end{center}
\vspace*{0cm}
\end{titlepage}
\vspace*{-0mm}
	
%%%%%%%%%%%%%%%%%%%%%%%%%%%%%%%%%%%%%%%%%%%%%%%%%%%%%%%%%%%%%%%%%%%%%%%%%%%%%%%%%%%%%%%%%%%%%%%%%%%%%%%%%%%%%%%%%%%%%%%%%%%%
	
\begin{abstract}
	In this paper, we generalize the covariant phase space formalism conjugate to field dependent vectors to include internal gauge transformations when gauge fields are present. In our formalism, the symmetry generators are combination of diffeomorphisms plus internal gauge transformations and depend on the field configuration. When the field dependence of the symmetry generators is considered in the covariant phase space formalism, the first law of black hole thermodynamics is a direct result for generally invariant gravitational theories. To check the validity of our formalism, we investigate the conserved charges and the first laws of thermodynamics for a torus-like black hole in Einstein-Maxwell theory and a charged Einstein-Euler-Heisenberg AdS black hole in Einstein-Euler-Heisenberg nonlinear electrodynamics.    
\end{abstract}

\newpage

\section{Introduction}
Symmetry plays an important role in modern physics. According to the well-known Noether’s theorem, symmetries are related to conservation laws \cite{Noether:1918zz}. The study of symmetry has become a cornerstone of modern physics and has provided a solid foundation for understanding the conservation laws in classical mechanics, general relativity and quantum field theory. In the realm of black hole physics, when we obtain an exact black hole solution, the first step is to determine the conserved charges, such as the mass, angular momentum and electric charge. In the early 1970s, Hawking’s classical area theorem states that the area of a black hole event horizon can never decrease \cite{Hawking:1971vc}. By considering semiclassical quantum effects in curved spacetime, Hawking found that a black hole can be regarded as a black body system with Hawking temperature $ T_{\mathrm{H}}$ \cite{Hawking:1974rv,Hawking:1975vcx}. On the other hand, Bekenstein postulated that the area of the event horizon could be regarded as the entropy of a black hole, referred to as the Bekenstein-Hawking entropy \cite{Bekenstein:1972tm,Bekenstein:1973ur}. Following the definitions of the relevant variables, the four laws of black hole thermodynamics were established by Bardeen, Carter and Hawking \cite{Bardeen:1973gs}. Since then, black hole as a thermodynamic system has been extensively studied. The most profound law of black hole thermodynamics is the first law, which relates the conserved charge variations to the chemical potentials of a black hole. To date, several approaches have been proposed to calculate the conserved charges and derive the first law of black hole thermodynamics, each with its own merits and demerits. 

In the well-known ADM approach \cite{Arnowitt:1962hi}, the conserved charges can be calculated for an asymptotically flat spacetime at the asymptotic infinity. But this approach fails for anti-de Sitter (AdS) spacetime. An extension to asymptotically AdS geometry was given by Abbott, Deser and Tekin (ADT) \cite{Abbott:1981ff,Abbott:1982jh,Deser:2002rt,Deser:2002jk,Senturk:2012yi} in a covariant manner. In this traditional ADT method, one needs to linearize the dynamical fields and use the equations of motion (EOM), i.e., on-shell. An off-shell generalization of the ADT method is given in Ref. \cite{Kim:2013zha}, and is called the off-shell ADT method or the quasi-local conserved charge formalism. Without resorting to a linearization of the dynamical fields, the Brown-York method is put forwarded in Ref. \cite{Brown:1992br} by introducing an appropriate counterterm into the action. This formulation is especially useful in the context of the AdS/CFT correspondence. But usually, it is difficult to find an appropriate counterterm, and this method is not covariant.

Based on the Noether procedure, the covariant phase space formalism (CPSF) was proposed in \cite{Lee:1990nz,Wald:1993nt,Iyer:1994ys} by Wald and Iyer (also called Wald’s formalism). For reviews of the CPSF, see Refs. \cite{Compere:2018aar, Grumiller:2022qhx}. In the framework of the CPSF, a convenient method for calculating conserved charges conjugate to `exact symmetries' of black hole solutions while incorporating the parametric variation was developed by K. Hajian \textit{et al}. and is called solution phase space method (SPSM) \cite{Hajian:2015xlp}. In the SPSM, the calculations of the conserved charges are independent of the choice of horizon or asymptotics. Along another line, a general theory of conserved charges based on the cohomology principles was developed by Barnich-Brandt and Comp{\`e}re (BBC) \cite{Barnich:2001jy,Barnich:2003xg,Barnich:2004uw,Barnich:2007bf}. These formalisms have been extensively applied to investigate black hole thermodynamic properties, and have been used particularly to derive the first law of black hole thermodynamics for generic theories of gravity.

In the original Wald’s formalism \cite{Lee:1990nz,Wald:1993nt,Iyer:1994ys}, the black hole entropy was defined as a Noether conserved charge associated with the normalized horizon Killing vectors, and was calculated by an integral over the bifurcation surface of a black hole event horizon. The first law of thermodynamics was then derived for a diffeomorphism invariant gravitational theory. In many cases, if gauge fields are present, as in Einstein-Maxwell and Einstein-Yang-Mills theories, one needs to consider a specific internal gauge transformation, which leads to the corresponding conserved charges. In these cases the symmetry generators $\epsilon$ are composed of spacetime diffeomorphism $\xi$ and internal gauge transformation $\lambda$, i.e., $\epsilon=(\xi, \lambda )$. Extensions of the CPSF that include internal gauge transformations are presented in Refs. \cite{Hajian:2015xlp,Prabhu:2015vua}. In the framework of the CPSF, the conserved charges conjugate to field dependent vectors $\xi=\xi (\Phi)$ are derived in Ref. \cite{Compere:2015knw}, Field dependent diffeomorphisms have been considered earlier in Refs. \cite{Gao:2003ys,Frodden:2017qwh,Chandrasekaran:2021vyu,Golshani:2026lvc}, and used to formulate the covariant phase space with fluctuating boundaries \cite{Golshani:2024fry} and the gravitational entropy \cite{Choi:2025dwa}. The field dependent vectors are also appeared in the form of modified Lie bracket \cite{Barnich:2001jy,Barnich:2011mi}.

In general, the symmetry generator $\epsilon$ is a function of the dynamical fields $\Phi$ or the solution parameters $p_{\alpha}$, i.e., $\epsilon=\epsilon (\Phi)$ and $\delta \epsilon \neq 0$, which includes $\delta \xi \neq 0$ and $\delta \lambda \neq 0$. In this paper, we generalize the CPSF conjugate to field dependent vectors \cite{Compere:2015knw} to include internal gauge transformations and consider $\delta \epsilon \neq 0$. We present a systematic derivation of the CPSF conjugate to field dependent symmetry generators. When the field dependence is considered in the CPSF, we prove that the first law of black hole thermodynamics is a direct result for generally invariant gravitational theories (up to total derivative terms). In the field dependent CPSF, the calculation of the conserved charges and the proof of the first law are independent of the choice of surface, or enter into a specifically gravitational theory. This is an advantage of the field dependent CPSF compared with other derivations of the first law in Refs. \cite{Gao:2003ys,Compere:2007vx,Jiang:2021pna}. To check the validity of our formalism, we investigate the conserved charges and the first laws of thermodynamics for a torus-like black hole in Einstein-Maxwell theory \cite{Huang:1995zb} and a charged Einstein-Euler-Heisenberg (EEH) AdS black hole in Einstein-Euler-Heisenberg nonlinear electrodynamics (NED) \cite{Magos:2020ykt}.

The rest of this paper is organized as follows. In Sec.2, we provide a systematic derivation of the CPSF conjugate to field dependent symmetry generators. In Sec.3, we use the CPSF conjugate to field dependent symmetry generators to prove the first law of black hole thermodynamics. In Sec.4, we investigate the conserved charges and the first laws of thermodynamics for a torus-like black hole in Einstein-Maxwell theory and a charged EEH-AdS black hole in EEH NED. Finally, we present our conclusions and outlook in Sec.5.

%%%%%%%%%%%%%%%%%%%%%%%%%%%%%%%%%%%%%%%%%%%%%%%%%%%%%%%%%%%%%%%%%%%%%%%%%%%%%%%%%%%%%%%%%%%%%%%%%%%%%%%%%%%%%%%%%%%%%%%%%%%%%%%%%%%%%%%%%%%%%%%%%%%%%%%%%%%%%%%%%%%%%%%%%%%%%%%%%%%%%%%%%%%%%%%%%%	
\section{Covariant phase space formalism conjugated to field dependent symmetry generators}
In this section, we generalize the CPSF conjugate to field dependent vectors \cite{Compere:2015knw} to include internal gauge transformations. When the gauge fields $\mathbf{A}^a$ are present in theories, for example, Einstein-Maxwell theory contains the $\mathrm{U(1)}$ gauge field $1$-form $\mathbf{A}$. Except to the spacetime diffeomorphism generated by a vector field $\xi$, the Lagrangian remains gauge invariant under the internal gauge transformation $\mathbf{A} \rightarrow \mathbf{A}+\mathrm{d} \lambda$ for an arbitrary scalar $\lambda$. Consequently, the symmetry generator $\epsilon$ is a combination of diffeomorphism plus gauge transformation, given by $\epsilon \equiv \{ \xi, \lambda \}$, and the general gauge transformation takes the form $\delta_{\epsilon} \Phi = \left\{ \mathscr{L}_{\xi} \Phi , \delta_{\lambda} \Phi \right\}$, where $\mathscr{L}_{\xi}$ denotes the Lie derivative along the smooth vector field $\xi$. In general, the generator $\epsilon$ is a function of the dynamical fields $\Phi$, i.e., $\epsilon=\epsilon (\Phi)$, and $\delta \epsilon \neq 0$ (which includes $\delta \xi \neq 0$, $\delta \lambda \neq 0 $).

For generality, we consider a generally invariant gravitational theory up to total derivative terms, for example, the gravitational Chern-Simons theory \cite{Tachikawa:2006sz}
\begin{equation}\label{gcs}
	\mathbf{L}_{\mathrm{CS}} \sim \mathrm{tr}\mathbf{\Gamma}\wedge\mathbf{R}^{2m+1}+\cdots ,
\end{equation}
or the matter Chern-Simons theory containing the term \cite{Compere:2009dp}
\begin{equation}\label{mcs}
	\mathbf{C}_{n}(\mathbf{A}) \sim \mathbf{A} \wedge \mathbf{F} \wedge \cdots \wedge \mathbf{F},
\end{equation}
where $\mathbf{\Gamma}$ is the affine connection, $\mathbf {R}$ is the curvature $2$-form, $\mathbf{F} = \mathrm{d} \mathbf{A} $ is the electromagnetic field strength associated with the gauge field $1$-form $\mathbf{A}$.

\textbf{Covariant phase space formalism:} A phase space is a manifold $\mathcal{M}$ equipped with a symplectic $2$-form $\Omega$. We consider an $n$-dimensional generally invariant gravitational theory up to total derivative terms described by the action
\begin{equation}
	\mathcal{S}(\Phi)=\int \mathbf{L}(\Phi),
\end{equation} 
where $\mathbf{L}$ is the Lagrangian $n$-form, and $\Phi$ collectively denotes all the dynamical fields, $\Phi = \left\{ g_{\mu \nu}, A_{\mu}, \cdots \right\}$. The variation of  $\mathbf{L}$ leads to
\begin{equation}\label{variationL}
	\delta \mathbf{L}(\Phi)=\mathbf{E}_{\Phi} \delta \Phi+\mathrm{d} \mathbf{\Theta}(\delta \Phi, \Phi) ,
\end{equation}
where $\delta \Phi$ is a generic field variation and provides a basis for the tangent bundle of the phase space $\mathcal{M}$. The set $\mathbf{E}_{\Phi}=0$ gives the equations of motion (EOM). $\mathbf{\Theta}$ is an $(n-1;1)$-form called the Lee-Wald symplectic potential, where the $(n-1)$-form is over the spacetime and the $1$-form is on the tangent bundle of $\mathcal{M}$. The Lee-Wald symplectic $2$-form in the CPSF is defined as \cite{Lee:1990nz}
\begin{equation}\label{key}
	\Omega \left(\delta_{1} \Phi, \delta_{2} \Phi, \Phi\right)=\int_{\Sigma} \bm{\omega} \left(\delta_{1} \Phi, \delta_{2} \Phi, \Phi\right) ,
\end{equation}
where
\begin{equation}\label{key}
	\bm{\omega} \left(\delta_{1} \Phi, \delta_{2} \Phi, \Phi \right) \equiv \delta_{1} \mathbf{\Theta} \left(\delta_{2} \Phi, \Phi\right)-\delta_{2} \mathbf{\Theta} \left(\delta_{1} \Phi, \Phi\right)
\end{equation} 
representing the symplectic current form, which is an $(n-1;2)$-form. $\Sigma$ is an arbitrary smooth codimension-$1$ surface. 

When $\Phi$ solves the EOM $\mathbf{E}_{\Phi}=0$ and $\delta \Phi$ solves the linearized EOM $\delta \mathbf{E}_{\Phi}=0$, i.e., on-shell, we have
\begin{equation}\label{key}
	\delta \mathbf{L}(\Phi) \approx \mathrm{d} \mathbf{\Theta}(\delta \Phi, \Phi) ,
\end{equation}
and the conservation condition
\begin{equation}\label{key}
	\mathrm{d} \bm{\omega} \left(\delta_{1} \Phi, \delta_{2} \Phi, \Phi\right) \approx 0
\end{equation}   	
is satisfied by $\left( \delta_{1} \delta_{2}-\delta_{2} \delta_{1} \right) \mathbf{L} (\Phi) \approx 0$.  Here and in what follows $ \approx$ denotes the on-shell equality. If we denote the set of dynamical fields by $\mathscr{F}$, i.e., $\Phi \in \mathscr{F}$, $(\mathscr{F},\Omega)$ constitutes a well-defined phase space. The tangent space of $\mathscr{F}$ is denoted by $T_{\mathscr{F}}$, $\delta \Phi \in T_{\mathscr{F}}$.

The variation of the Lagrangian $\mathbf{L}$ induced by $\epsilon$ is gauge invariant up to a total derivative term
\begin{equation}\label{deM}
	\delta_{\epsilon}\mathbf{L}(\Phi)=\mathrm{d}\mathbf{M}_{\epsilon}(\Phi) ,
\end{equation} 
where $\delta_{\epsilon}\mathbf{L}(\Phi)=\delta_{\xi}\mathbf{L}(\Phi)+\delta_{\lambda}\mathbf{L}(\Phi)=\mathscr{L}_{\xi}\mathbf{L}(\Phi)+\mathrm{d} \mathbf{\Xi}_{\xi}(\Phi)+\mathrm{d}(\lambda \mathrm{d}\mathbf{C}_{n-2} (\mathbf{A}) )$, the term $\mathbf{\Xi}_{\xi}(\Phi)$ comes from the gravitational Chern-Simons term \cite{Tachikawa:2006sz}, and $\lambda \mathrm{d}\mathbf{C}_{n-2} (\mathbf{A})$ comes from the matter Chern-Simons term \cite{Compere:2009dp}. According to the identity $\mathscr{L}_{\xi}\mathbf{L}(\Phi)=\xi \cdot \mathrm{d}\mathbf{L}(\Phi)+\mathrm{d}(\xi \cdot \mathbf{L}(\Phi))$, and $\mathrm{d}\mathbf{L}(\Phi)=0$ (for a top form $\mathbf{L}(\Phi)$), the total derivative term is given by
\begin{equation}\label{key}
	\mathbf{M}_{\epsilon}(\Phi)=\xi \cdot \mathbf{L}(\Phi)+\mathbf{\Xi}_{\xi}(\Phi)+\lambda \mathrm{d}\mathbf{C}_{n-2} (\mathbf{A}) .
\end{equation}
On the other hand, the variation of the Lagrangian $\mathbf{L}$ under the transformation generated by $\epsilon$ is
\begin{equation}\label{deL}
	\delta_{\epsilon} \mathbf{L}(\Phi)=\mathbf{E}_{\Phi} \delta_{\epsilon} \Phi+\mathrm{d} \mathbf{\Theta}(\delta_{\epsilon} \Phi, \Phi) .
\end{equation}
From Eq.(\ref{deM}) and Eq.(\ref{deL}), we obtain
\begin{equation}\label{key}
	-\mathbf{E}_{\Phi} \delta_{\epsilon} \Phi=\mathrm{d}\left( \mathbf{\Theta}(\delta_{\epsilon} \Phi, \Phi)-\mathbf{M}_{\epsilon}(\Phi) \right) .
\end{equation}
We define the standard Noether $(n-1)$-form current $\mathbf{J}_{\epsilon}(\Phi)$ as
\begin{equation}\label{Noether current}
	\mathbf{J}_{\epsilon}(\Phi)=\mathbf{\Theta}(\delta_{\epsilon} \Phi, \Phi)-\mathbf{M}_{\epsilon}(\Phi) ,
\end{equation}
which satisfies $\mathrm{d}\mathbf{J}_{\epsilon}(\Phi) \approx 0$. According to Poincar{\'e}’s Lemma, since $\mathbf{J}_{\epsilon}(\Phi)$ is closed, it is locally exact on-shell, and hence there exists an $(n-2)$-form $\mathbf{Q}_{\epsilon}(\Phi)$, satisfying
\begin{equation}\label{Ncq}
	\mathbf{J}_{\epsilon}(\Phi) \approx \mathrm{d} \mathbf{Q}_{\epsilon}(\Phi),
\end{equation}
where $\mathbf{Q}_{\epsilon}(\Phi)$ is called the Noether-Wald charge density. For $\delta \epsilon \neq 0$, the variation of $\mathbf{M}_{\epsilon}(\Phi)$ is given by
\begin{align}\label{varM}
	\delta\mathbf{M}_{\epsilon}(\Phi)&=\delta\xi \cdot \mathbf{L}(\Phi)+\xi \cdot \delta\mathbf{L}(\Phi)+\delta\mathbf{\Xi}_{\xi}(\Phi)+\delta\lambda \mathrm{d}\mathbf{C}_{n-2} (\mathbf{A})+\lambda \mathrm{d}\delta\mathbf{C}_{n-2} (\mathbf{A}) \nonumber \\
	&\approx \mathscr{L}_{\xi}\mathbf{\Theta}(\delta \Phi, \Phi)+\mathrm{d}(-\xi \cdot \mathbf{\Theta}(\delta \Phi, \Phi))+\delta\xi \cdot \mathbf{L}(\Phi)+\delta\mathbf{\Xi}_{\xi}(\Phi)+\delta\lambda \mathrm{d}\mathbf{C}_{n-2} (\mathbf{A})+\lambda \mathrm{d}\delta\mathbf{C}_{n-2} (\mathbf{A}) \nonumber \\
	&=\mathscr{L}_{\xi}\mathbf{\Theta}(\delta \Phi, \Phi)+\mathrm{d}(-\xi \cdot \mathbf{\Theta}(\delta \Phi, \Phi))+\mathbf{M}_{\delta \epsilon}(\Phi)+\delta^{[\Phi]} \mathbf{\Xi}_{\xi}(\Phi)+\lambda \mathrm{d}\delta\mathbf{C}_{n-2} (\mathbf{A}) ,
\end{align}
where
\begin{equation}\label{key}
	\mathbf{M}_{\delta \epsilon}(\Phi)=\delta\xi \cdot \mathbf{L}(\Phi)+\mathbf{\Xi}_{\delta \xi}(\Phi)+\delta\lambda \mathrm{d}\mathbf{C}_{n-2} (\mathbf{A}) ,
\end{equation}
and $\delta^{[ \Phi ]}$ denotes that $\delta$ acts only on the explicit field dependence $\Phi$, but not on the $\xi$ inside $\mathbf{\Xi}_{\xi}(\Phi)$. From Eq.(\ref{varM}), we have
\begin{equation}\label{doM}
	\delta^{[\Phi]}\mathbf{M}_{\epsilon}(\Phi)=\mathscr{L}_{\xi}\mathbf{\Theta}(\delta \Phi, \Phi)+\mathrm{d}(-\xi \cdot \mathbf{\Theta}(\delta \Phi, \Phi))+\delta^{[\Phi]} \mathbf{\Xi}_{\xi}(\Phi)+\lambda \mathrm{d}\delta\mathbf{C}_{n-2} (\mathbf{A}) .
\end{equation}
As in Ref. \cite{Compere:2009dp}, we define $\mathbf{\Pi}_{\epsilon}(\delta \Phi, \Phi)$ through the equation
\begin{equation}\label{vartheta}
	\delta_{\epsilon} \mathbf{\Theta}(\delta \Phi, \Phi)=\mathscr{L}_{\xi}\mathbf{\Theta}(\delta \Phi, \Phi)+\mathbf{\Pi}_{\epsilon}(\delta \Phi, \Phi) .
\end{equation} 
We calculate $\delta \delta_{\epsilon}\mathbf{L}(\Phi)$ in two ways by taking the gauge transformation $\delta_{\epsilon}$ of Eq. (\ref{variationL}) and the variation $\delta^{[\Phi]}$ of Eq. (\ref{deM}) as
\begin{align}\label{key}
	0 &=\delta^{[\Phi]} \delta_{\epsilon}\mathbf{L}(\Phi)-\delta_{\epsilon} \delta^{[\Phi]}\mathbf{L}(\Phi) \nonumber \\
	&\approx \mathrm{d}(\delta^{[\Phi]}\mathbf{M}_{\epsilon}(\Phi)-\delta_{\epsilon}\mathbf{\Theta}(\delta \Phi, \Phi) ) \nonumber \\
	&\approx \mathrm{d}(\delta^{[\Phi]} \mathbf{\Xi}_{\xi}(\Phi)+\lambda \mathrm{d}\delta\mathbf{C}_{n-2} (\mathbf{A})-\mathbf{\Pi}_{\epsilon}(\delta \Phi, \Phi) ),
\end{align}
in the last equality, where Eqs.(\ref{doM}) and (\ref{vartheta}) are used.
Therefore, locally, there exists an $(n-2)$-form $\mathbf{\Sigma}_{\epsilon}(\delta \Phi, \Phi)$ such that
\begin{equation}\label{dSigma}
	\delta^{[\Phi]} \mathbf{\Xi}_{\xi}(\Phi)+\lambda \mathrm{d}\delta\mathbf{C}_{n-2} (\mathbf{A})-\mathbf{\Pi}_{\epsilon}(\delta \Phi, \Phi) \approx \mathrm{d}\mathbf{\Sigma}_{\epsilon}(\delta \Phi, \Phi) .
\end{equation}
Thus, substituting Eqs.(\ref{vartheta}) and (\ref{dSigma}) into Eq. (\ref{varM}), we have
\begin{align}\label{key}
	\delta\mathbf{M}_{\epsilon}(\Phi)&=\delta_{\epsilon}\mathbf{\Theta}(\delta \Phi, \Phi)+\mathbf{M}_{\delta \epsilon}(\Phi)+\mathrm{d}(-\xi \cdot \mathbf{\Theta}(\delta \Phi, \Phi))+\delta^{[\Phi]} \mathbf{\Xi}_{\xi}(\Phi)+\lambda \mathrm{d}\delta\mathbf{C}_{n-2} (\mathbf{A})-\mathbf{\Pi}_{\epsilon}(\delta \Phi, \Phi) \nonumber \\
	&=\delta_{\epsilon}\mathbf{\Theta}(\delta \Phi, \Phi)+\mathbf{M}_{\delta \epsilon}(\Phi)+\mathrm{d}(-\xi \cdot \mathbf{\Theta}(\delta \Phi, \Phi)+\mathbf{\Sigma}_{\epsilon}(\delta \Phi, \Phi)) .
\end{align} 
The variation of the Noether current $\mathbf{J}_{\epsilon}(\Phi)$ in Eq.(\ref{Noether current}) is given by
\begin{align}\label{key}
	\delta \mathbf{J}_{\epsilon}(\Phi)&=\delta \mathbf{\Theta}(\delta_{\epsilon} \Phi, \Phi)-\delta \mathbf{M}_{\epsilon}(\Phi) \nonumber \\
	&=\delta^{[\Phi]} \mathbf{\Theta}(\delta_{\epsilon} \Phi, \Phi)-\delta_{\epsilon}\mathbf{\Theta}(\delta \Phi, \Phi)+\mathbf{\Theta}(\delta_{\delta \epsilon} \Phi, \Phi)-\mathbf{M}_{\delta \epsilon}(\Phi)-\mathrm{d}(-\xi \cdot \mathbf{\Theta}(\delta \Phi, \Phi)+\mathbf{\Sigma}_{\epsilon}(\delta \Phi, \Phi) ) \nonumber \\
	&=\bm{\omega} \left(\delta \Phi, \delta_{\epsilon} \Phi, \Phi \right)+\mathbf{J}_{\delta \epsilon}(\Phi)-\mathrm{d}(-\xi \cdot \mathbf{\Theta}(\delta \Phi, \Phi)+\mathbf{\Sigma}_{\epsilon}(\delta \Phi, \Phi) ) .
\end{align}
where
\begin{equation}\label{key}
	\delta \mathbf{\Theta}(\delta_{\epsilon} \Phi, \Phi)=\delta^{[\Phi]} \mathbf{\Theta}(\delta_{\epsilon} \Phi, \Phi)+\mathbf{\Theta}(\delta_{\delta \epsilon} \Phi, \Phi),
\end{equation}
\begin{equation}\label{key}
	\mathbf{J}_{\delta \epsilon}(\Phi)=\mathbf{\Theta}(\delta_{\delta \epsilon} \Phi, \Phi)-\mathbf{M}_{\delta \epsilon}(\Phi),
\end{equation}
and the Lee-Wald symplectic current form is defined by
\begin{equation}\label{key}
	\bm{\omega} \left(\delta \Phi, \delta_{\epsilon} \Phi, \Phi \right)=\delta^{[\Phi]} \mathbf{\Theta}(\delta_{\epsilon} \Phi, \Phi)-\delta_{\epsilon}\mathbf{\Theta}(\delta \Phi, \Phi) .
\end{equation}
Reordering the terms and using $\mathbf{J}_{\epsilon}(\Phi) \approx \mathrm{d} \mathbf{Q}_{\epsilon}(\Phi)$, we get
\begin{align}\label{key}
	\bm{\omega} \left(\delta \Phi, \delta_{\epsilon} \Phi, \Phi \right)&\approx \mathrm{d}(\delta \mathbf{Q}_{\epsilon}(\Phi)-\mathbf{Q}_{\delta \epsilon}(\Phi) -\xi \cdot \mathbf{\Theta}(\delta \Phi, \Phi)+\mathbf{\Sigma}_{\epsilon}(\delta \Phi, \Phi) ) \nonumber \\
	&=\mathrm{d} \bm{k}_{\epsilon}(\delta \Phi, \Phi) ,
\end{align}
with
\begin{align}\label{k expression}
	\bm{k}_{\epsilon}(\delta \Phi, \Phi)&=\delta \mathbf{Q}_{\epsilon}(\Phi)-\mathbf{Q}_{\delta \epsilon}(\Phi) -\xi \cdot \mathbf{\Theta}(\delta \Phi, \Phi)+\mathbf{\Sigma}_{\epsilon}(\delta \Phi, \Phi) \nonumber \\
	&=\delta^{[\Phi]} \mathbf{Q}_{\epsilon}(\Phi) -\xi \cdot \mathbf{\Theta}(\delta \Phi, \Phi)+\mathbf{\Sigma}_{\epsilon}(\delta \Phi, \Phi) .
\end{align}
Here $\bm{k}_{\epsilon}(\delta \Phi, \Phi)$ is an $(n-2, 1)$-form, called the surface charge density. Note that, for an exact gauge invariant theory, the term $\mathbf{\Sigma}_{\epsilon}(\delta \Phi, \Phi)$ vanishes. Therefore, the infinitesimal charge variation $\delta H_{\epsilon} (\Phi)$ conjugate to the generator $\epsilon$ over an arbitrary closed and smooth codimension-$2$ surface $\partial \Sigma$ can be defined as
\begin{align} \label{eqdh}
	\delta H_{\epsilon}(\Phi) &\equiv \Omega\left(\delta \Phi, \delta_{\epsilon} \Phi, \Phi\right)=\int_{\Sigma} \bm{\omega}\left(\delta \Phi, \delta_{\epsilon} \Phi, \Phi\right) \approx \oint_{\partial \Sigma} \bm{k}_{\epsilon}(\delta \Phi, \Phi) \nonumber\\
	&= \oint_{\partial \Sigma}(\delta^{[\Phi]} \mathbf{Q}_{\epsilon}(\Phi) -\xi \cdot \mathbf{\Theta}(\delta \Phi, \Phi)+\mathbf{\Sigma}_{\epsilon}(\delta \Phi, \Phi)) \nonumber \\
	&= \oint_{\partial \Sigma}(\delta \mathbf{Q}_{\epsilon}(\Phi)-\mathbf{Q}_{\delta \epsilon}(\Phi) -\xi \cdot \mathbf{\Theta}(\delta \Phi, \Phi)+\mathbf{\Sigma}_{\epsilon}(\delta \Phi, \Phi)) .
\end{align}

If $\delta H_{\epsilon} (\Phi)$ is well-defined and integrable, we can determine the conserved charge $ H_{\epsilon} (\Phi)$. The integrability condition is basically
\begin{equation}\label{key}
	\left(\delta_{1} \delta_{2}-\delta_{2} \delta_{1}\right) H_{\epsilon}(\Phi)=0 ,
\end{equation}
which is equivalent to \cite{Hajian:2015xlp}
\begin{equation}\label{intcondition}
	\oint_{\partial \Sigma}\left(\xi \cdot \bm{\omega}\left(\delta_{1} \Phi, \delta_{2} \Phi, \Phi\right)+\bm{k}_{\delta_{1} \epsilon}\left(\delta_{2} \Phi, \Phi\right)-\bm{k}_{\delta_{2} \epsilon}\left(\delta_{1} \Phi, \Phi\right)\right) \approx 0 .
\end{equation}

To ensure the conservation of charge variation $\delta H_{\epsilon} (\Phi)$ and the independence of integration on $\Sigma$ and $\partial \Sigma$, the symplectic current form $\bm{\omega}$ must vanish on-shell for a certain subclass of $\delta_{\epsilon} \Phi$'s, i.e., 
\begin{equation}\label{symplectic symmetry}
	\bm{\omega}\left(\delta \Phi, \delta_{\epsilon} \Phi, \Phi\right) \approx 0 .
\end{equation}
In this case the transformations generated by $\delta_{\epsilon} \Phi$ are called symplectic symmetries. The family of $\epsilon$'s with this property can be divided into two sets \cite{Hajian:2015xlp}: (a) non-exact symmetry generators denoted by $\chi$, for which $\delta_{\chi} \Phi \neq 0$ at least at one point in the phase space; (b) exact symmetry generators denoted by $\eta$, for which $\delta_{\eta} \Phi = 0$ all over the phase space.

Under the symplectic symmetry condition in Eq.(\ref{symplectic symmetry}), we have
\begin{align}\label{key}
	0 &=\int_{\Sigma} \bm{\omega}\left(\delta \Phi, \delta_{\epsilon} \Phi, \Phi\right) \nonumber \\
	&=\oint_{\partial \Sigma_{2}} \bm{k}_{\epsilon}(\delta \Phi, \Phi)-\oint_{\partial \Sigma_{1}} \bm{k}_{\epsilon}(\delta \Phi, \Phi) ,
\end{align}
where $\partial \Sigma = \partial \Sigma_{2} \cup \partial \Sigma_{1} $. This ensures the conservation of $\delta H_{\epsilon} (\Phi)$ and thus the independence of integration on $\Sigma$ and $\partial \Sigma$, i.e., $\Sigma$ and $\partial \Sigma$ are arbitrary.

For solutions $\Phi (x^{\mu}, p_{\alpha})$ of the EOM $\mathbf{E}_{\Phi}=0$, where $p_{\alpha}$ is a set of solution parameters, if the integration in Eq.(\ref{eqdh}) is well-defined and integrable over the solution parameters $p_{\alpha}$, the conserved charges $H_{\epsilon} (p_{\alpha})$ could be calculated by
\begin{equation}\label{Hexpression}
	H_{\epsilon}( \Phi (p_{\alpha}),\bar{\Phi} (\bar{p}_{\alpha}) )=\int_{\bar{p}}^{p} \delta H_{\epsilon}(\Phi (p_{\alpha}))+H_{\epsilon} ( \bar{\Phi} (\bar{p}_{\alpha}) ) ,
\end{equation}
where the integration is performed over an arbitrary integral curve connecting a reference (or background) field configuration $\bar{\Phi} (x^{\mu}, \bar{p}_{\alpha})$ to the target space $\Phi (x^{\mu}, p_{\alpha})$ under considerations. $H_{\epsilon}(\bar{\Phi}(\bar{p}_{\alpha}))$ is the reference point for $ H_{\epsilon}$ defined on $\bar{\Phi} (x^{\mu}, \bar{p}_{\alpha})$.

%%%%%%%%%%%%%%%%%%%%%%%%%%%%%%%%%%%%%%%%%%%%%%%%%%%%%%%%%%%%%%%%%%%%%%%%%%%%%%%%%%%%%%%%%%%%%%%%%%%%%%%%%%%%%%%%%%%%%%%%%%%%%%%%%%%%%%%%%%%%%%%%%%%%%%%%%%%%%%%%%%%%%%%%%%%%%%%%%%%%%%%%%%%%%%%%%%	
\section{The first law}	
In this section we use the CPSF conjugate to field dependent symmetry generators to prove the first law of black hole thermodynamics for arbitrary generally invariant gravitational theories (up to total derivative terms). For generality, we consider stationary axisymmetric black hole solutions in theories that include some internal gauge transformations $\lambda^{a}$, when the gauge fields $\mathbf{A}^{a}$ are present. Let us denote the timelike Killing vector of the stationary black hole by $\partial_{t}$ and its possible other axial isometries by $\partial_{\varphi^{i}}$. Here we choose the coordinates $\{ t, \cdots, \varphi^{i} \}$, and assume that the conserved charge variations are well-defined and integrable. The symmetry generators $\epsilon$ to which the mass $M$ and angular momentum $J_{i}$ are attributed chosen as $\eta_{{}_M}=\{ \partial_{t}, 0 \}$ and $\eta_{{}_ {J_i}}=\{ -\partial_{\varphi^{i}}, 0 \}$, respectively. The gauge charge $Q_{a}$ associated with the gauge field $\mathbf{A}^{a}$ is conjugate to the global internal gauge transformation $\eta_{{}_{Q_a}}=\{ 0, -1^a \}$, in which $1^a$ means $\lambda^a=1$, and $\lambda^b=0$ for $b \neq a$. Therefore, the conserved charge variations are $\delta M \equiv \delta H_{\eta_{{}_{M}} }$, $\delta J_i \equiv \delta H_{\eta_{{}_{J_i}}}$, and $\delta Q_{a} \equiv \delta H_{\eta_{{}_{Q_a}}}$.

In the field dependent CPSF, we also define the entropy as the conserved charge $\delta S = \delta H_{\eta_{{}_{\mathrm{H}}} }$. Here $\eta_{{}_{\mathrm{H}}} $ is the symmetry generator conjugate to the entropy $S$, and is a combination of other symmetry generators, given by
\begin{align}\label{key}
	\eta_{{}_{\mathrm{H}}}&=\big\{ \frac{1}{T_{\mathrm{H}}}(\partial_{t}+\Omega^i_{\mathrm{H}} \partial_{\varphi^{i}} ), -\frac{1}{T_{\mathrm{H}}} \Phi^a_{\mathrm{H}} \big\} \nonumber \\
	&=\frac{1}{T_{\mathrm{H}}}\eta_{{}_M}-\frac{\Omega^i_{\mathrm{H}}}{T_{\mathrm{H}}} \eta_{{}_ {J_i}}-\frac{\Phi^a_{\mathrm{H}}}{T_{\mathrm{H}}} \eta_{{}_{Q_a}} ,
\end{align}
where, for each horizon, the temperature $T_{\mathrm{H}}= \frac{\kappa_{{}_{\mathrm{H}}}}{2\pi}$, angular velocities $\Omega^i_{\mathrm{H}}$, electric potentials $\Phi^a_{\mathrm{H}}$ and surface gravity $\kappa_{{}_{\mathrm{H}}}$ are defined. Although $\delta \eta_{{}_M} = 0$, $\delta \eta_{{}_ {J_i}} = 0$, and $\delta \eta_{{}_{Q_a}} = 0$, the chemical potentials $ \left(T_{\mathrm{H}}, \Omega^i_{\mathrm{H}}, \Phi^a_{\mathrm{H}}, \cdots \right) $ are constants on the spacetime but are functions of the solution parameters, i.e., $\delta T_{\mathrm{H}} \neq 0 $, $\delta \Omega^i_{\mathrm{H}} \neq 0 $, $\delta \Phi^a_{\mathrm{H}} \neq 0 $ etc. Thus, in general, the entropy generator satisfies $\delta \eta_{{}_{\mathrm{H}}} \neq 0$.

From the charge variation in Eq.(\ref{eqdh}) and the linearity of $\bm{k}$ in generators $\epsilon$, we have
\begin{align}\label{proof}
	\delta S = \delta H_{\eta_{{}_{\mathrm{H}}}}&= \oint_{\partial \Sigma}(\delta (\frac{1}{T_{\mathrm{H}}}\mathbf{Q}_{\eta_{{}_{M}}})-\delta(\frac{1}{T_{\mathrm{H}}})\mathbf{Q}_{\eta_{{}_{M}}}-\frac{1}{T_{\mathrm{H}}} \partial_{t}\cdot \mathbf{\Theta}(\delta \Phi, \Phi)-\frac{1}{T_{\mathrm{H}}} \mathbf{\Sigma}_{\eta_{{}_{M}}}(\delta \Phi, \Phi) ) \nonumber \\
	&-\oint_{\partial \Sigma}(\delta (\frac{\Omega^i_{\mathrm{H}}}{T_{\mathrm{H}}}\mathbf{Q}_{\eta_{{}_{J_i}}})-\delta(\frac{\Omega^i_{\mathrm{H}}}{T_{\mathrm{H}}})\mathbf{Q}_{\eta_{{}_{J_i}}}-\frac{\Omega^i_{\mathrm{H}}}{T_{\mathrm{H}}} (-\partial_{\varphi^{i}})\cdot \mathbf{\Theta}(\delta \Phi, \Phi)-\frac{\Omega^i_{\mathrm{H}}}{T_{\mathrm{H}}} \mathbf{\Sigma}_{\eta_{{}_{J_i}}}(\delta \Phi, \Phi) ) \nonumber \\
	&-\oint_{\partial \Sigma}(\delta (\frac{\Phi^a_{\mathrm{H}}}{T_{\mathrm{H}}}\mathbf{Q}_{\eta_{{}_{Q_a}}})-\delta(\frac{\Phi^a_{\mathrm{H}}}{T_{\mathrm{H}}})\mathbf{Q}_{\eta_{{}_{Q_a}}}-\frac{\Phi^a_{\mathrm{H}}}{T_{\mathrm{H}}} \mathbf{\Sigma}_{\eta_{{}_{Q_a}}}(\delta \Phi, \Phi) ) \nonumber \\
	&=\frac{1}{T_{\mathrm{H}}} \oint_{\partial \Sigma}(\delta \mathbf{Q}_{\eta_{{}_{M}}}- \partial_{t}\cdot \mathbf{\Theta}(\delta \Phi, \Phi)- \mathbf{\Sigma}_{ \{\partial_{t},0 \} }(\delta \Phi, \Phi) ) \nonumber \\
	&-\frac{\Omega^i_{\mathrm{H}}}{T_{\mathrm{H}}} \oint_{\partial \Sigma}(\delta \mathbf{Q}_{\eta_{{}_{J_i}}}-(-\partial_{\varphi^{i}})\cdot \mathbf{\Theta}(\delta \Phi, \Phi)-\mathbf{\Sigma}_{ \{ -\partial_{ \varphi^{i} }, 0 \} }(\delta \Phi, \Phi) ) \nonumber \\
	&-\frac{\Phi^a_{\mathrm{H}}}{T_{\mathrm{H}}} \oint_{\partial \Sigma}(\delta \mathbf{Q}_{\eta_{{}_{Q_a}}}-\mathbf{\Sigma}_{ \{0,1^a \} }(\delta \Phi, \Phi) ) \nonumber \\
	&=\frac{1}{T_{\mathrm{H}}} \delta M-\frac{\Omega^i_{\mathrm{H}}}{T_{\mathrm{H}}} \delta J_i-\frac{\Phi^a_{\mathrm{H}}}{T_{\mathrm{H}}} \delta Q_{a} ,
\end{align}
where $\mathbf{Q}_{\frac{1}{T_{\mathrm{H}}} \eta_{{}_{M}}}=\frac{1}{T_{\mathrm{H}}}\mathbf{Q}_{\eta_{{}_{M}}}$, $\mathbf{Q}_{\delta(\frac{1}{T_{\mathrm{H}}} \eta_{{}_{M}})}=\delta(\frac{1}{T_{\mathrm{H}}})\mathbf{Q}_{\eta_{{}_{M}}}$ etc, and
\begin{align}\label{key}
	\delta M &=\delta H_{\eta_{{}_{M}} }=\oint_{\partial \Sigma}(\delta \mathbf{Q}_{\eta_{{}_{M}}}- \partial_{t}\cdot \mathbf{\Theta}(\delta \Phi, \Phi)- \mathbf{\Sigma}_{ \{\partial_{t},0 \} }(\delta \Phi, \Phi) ) ,   \nonumber \\
	\delta J_i &=\delta H_{\eta_{{}_{J_i}}}=\oint_{\partial \Sigma}(\delta \mathbf{Q}_{\eta_{{}_{J_i}}}-(-\partial_{\varphi^{i}})\cdot \mathbf{\Theta}(\delta \Phi, \Phi)-\mathbf{\Sigma}_{ \{ -\partial_{ \varphi^{i} }, 0 \} }(\delta \Phi, \Phi) ) ,     \nonumber \\
	\delta Q_{a} &=\delta H_{\eta_{{}_{Q_a}}}=\oint_{\partial \Sigma}(\delta \mathbf{Q}_{\eta_{{}_{Q_a}}}-\mathbf{\Sigma}_{ \{0,1^a \} }(\delta \Phi, \Phi) ) ,   
\end{align}
the first law of black hole thermodynamics is proved:
\begin{equation}\label{key}
	\delta M=T_{\mathrm{H}}\delta S+\Omega^i_{\mathrm{H}}\delta J_i+\Phi^a_{\mathrm{H}}\delta Q_{a} .
\end{equation}

In our proof of the first law, the integration surface $\partial \Sigma$ is an arbitrary smooth, closed, and compact codimension-$2$ surface surrounding the black hole singularity. We are not limited to evaluating the integrals at the bifurcate Killing horizon or in the asymptotic region of spacetime. Only the chemical potentials $ \left(T_{\mathrm{H}}, \Omega^i_{\mathrm{H}}, \Phi^a_{\mathrm{H}}, \cdots \right) $ need to be calculated at the horizons. When the field dependence of the symmetry generators is considered in the CPSF, the first law of black hole thermodynamics is a direct result for generally invariant gravitational theories (up to total derivative terms).
%     
%    It is straightforward to generalize our formalism to theories which include the gauge field $(n-1)$-form $\mathbf{A}^{a}$ and with gauge symmetries $\mathbf{A}^{a} \rightarrow \mathbf{A}^{a}+\mathrm{d} \bm{\lambda}^{a}$,where $\bm{\lambda}^{a}$ is an $(n-2)$-form gauge symmetry generator []. 
For generic black hole solutions, such as Kerr-Newman (AdS) black holes, the generators conjugate to mass, angular momentum and electric charges are independent of the fields or solution parameters. Thus the conserved charges and the first law agree with those obtained using other methods, and we do not repeat these procedures. To illustrate the validity of our formalism, in the next section we will investigate the conserved charges and the first laws of thermodynamics for a torus-like black hole in Einstein-Maxwell theory \cite{Huang:1995zb} and a charged EEH-AdS black hole in EEH NED \cite{Magos:2020ykt}. They are sufficient for our purpose.

%%%%%%%%%%%%%%%%%%%%%%%%%%%%%%%%%%%%%%%%%%%%%%%%%%%%%%%%%%%%%%%%%%%%%%%%%%%%%%%%%%%%%%%%%%%%%%%%%%%%%%%%%%%%%%%%%%%%%%%%%%%%%%%%%%%%%%%%%%%%%%%%%%%%%%%%%%%%%%%%%%%%%%%%%%%%%%%%%%%%%%%%%%%%%%%%%%	
\section{Examples}
\subsection{Einstein-Maxwell theory}
We consider Einstein-Maxwell theory with cosmological constant $\Lambda$. The dynamical fields $\Phi$ are the metric $g_{\mu \nu}$ and the electromagnetic gauge field $A_{\mu}$ governed by the Lagrangian
\begin{equation}\label{key}
	L=\frac{1}{16 \pi}\left(R-2\Lambda-F_{\mu \nu}F^{\mu \nu} \right),
\end{equation}
where $R$ is the Ricci scalar, $F_{\mu \nu}=\nabla_{\mu} A_{\nu}-\nabla_{\nu} A_{\mu}$ is the electromagnetic field strength, and we set the Newtonian constant $G=1$.

The Lagrangian $n$-form is the Hodge dual of $L$,
\begin{equation}\label{key}
	\mathbf{L}=\star{L}=L\, \bm{\epsilon} ,
\end{equation}
where $ \bm{\epsilon}=\sqrt{-g}\,d^nx=\frac{\sqrt{-g}}{n!} \varepsilon_{\mu_{1} \cdots\mu_{n}} d x^{\mu_{1}} \wedge \cdots \wedge d x^{\mu_{n}} $ is the $n$-dimensional volume form, $g$ is the determinant of the metric $g_{\mu \nu}$, and $\varepsilon_{\mu_{1} \cdots\mu_{n}}$ is the Levi-Civita symbol.
The variations of the Lagrangian $n$-form $\mathbf{L}$ with respect to $g_{\mu \nu}$ and $A_{\mu}$ are
\begin{equation}\label{key}
	\delta \mathbf{L}(\Phi)=\left( \mathrm{E}^{\mu \nu}_{g} \delta g_{\mu \nu}+\mathrm{E}^{\mu}_{A} \delta A_{\mu} \right)\bm{\epsilon} +\mathrm{d} \mathbf{\Theta}(\delta \Phi, \Phi) ,
\end{equation}
where the EOMs are
\begin{align}\label{EOMs}
	\mathrm{E}^{\mu \nu}_{g} &=\frac{1}{16 \pi} \left[-(G^{\mu \nu}+\Lambda g^{\mu \nu})+2(F^{\mu \sigma}F^{\nu}{}_{\sigma}-\frac{1}{4}g^{\mu \nu}F_{\alpha \beta}F^{\alpha \beta} ) \right], \nonumber \\
	\mathrm{E}^{\mu}_{A} &=\frac{1}{16 \pi} \nabla_{\nu} \left[ -4 F^{\mu \nu} \right] ,
\end{align}
with
\begin{equation}\label{key}
	G_{\mu \nu} =R_{\mu \nu}-\frac{1}{2} R g_{\mu \nu}.
\end{equation}
The symplectic potential $\mathbf{\Theta}(\delta \Phi, \Phi)$ is given by
\begin{equation}\label{key}
	\mathbf{\Theta}(\delta \Phi, \Phi)=\Theta^{\mu}(\delta \Phi, \Phi) \sqrt{-g}(d^{n-1}x)_{\mu} ,
\end{equation}
where
\begin{equation}\label{key}
	\Theta^{\mu}(\delta \Phi, \Phi)=\frac{1}{16 \pi} \left[ 2 \nabla^{[\nu } {h^{\mu ]}}_{\nu}-4 F^{\mu \nu} \delta A_{\nu} \right] .
\end{equation}
Here and in what follows $h_{\mu \nu }=\delta g_{\mu \nu },\,\,\,\,h^{\mu \nu }= g^{\mu \alpha } g^{\nu \beta }\delta g_{\alpha \beta }=-\delta g^{\mu \nu },\,\,\,\,h=g^{\mu \nu }\delta g_{\mu \nu }$, and $\left(d^{n-p} x\right)_{\mu_{1} \cdots \mu_{p}}=\frac{1}{p !(n-p) !} \varepsilon_{\mu_{1} \cdots \mu_{p} \nu_{p+1} \cdots \nu_{n}} d x^{\nu_{p+1}} \wedge \cdots \wedge d x^{\nu_{n}}$.

For the generator $\epsilon=(\xi, \lambda)$, the field transformations are
\begin{align}\label{pvariations}
	\delta_{\epsilon}g_{\mu \nu}&=\mathscr{L}_{\xi}g_{\mu \nu}=\nabla_{\mu}\xi_{\nu}+\nabla_{\nu}\xi_{\mu}, \nonumber \\
	\delta_{\epsilon}A_{\mu}&=\mathscr{L}_{\xi}A_{\mu}+\nabla_{\mu}\lambda=\xi^{\sigma}F_{\sigma \mu}+\nabla_{\mu}\left( A_{\sigma}\xi^{\sigma}+\lambda \right),
\end{align}
From Eq.(\ref{Ncq}) we can get the Noether-Wald charge density
\begin{equation}\label{key}
	\mathbf{Q}_{\epsilon}(\Phi)=\mathrm{Q}^{\mu \nu}_{\epsilon}(\Phi) \sqrt{-g}(d^{n-2}x)_{\mu \nu} ,
\end{equation}
where
\begin{equation}\label{key}
	\mathrm{Q}^{\mu \nu}_{\epsilon}(\Phi)=\frac{1}{16 \pi}\left[-2 \nabla^{[\mu} \xi^{\nu ] } -4 F^{\mu \nu} (A_{\sigma}\xi^{\sigma}+\lambda) \right].
\end{equation}
Considering $\delta \epsilon \neq 0$, the variation of the Noether-Wald charge density is given by 
\begin{align}\label{key}
\delta( \sqrt{-g} \mathrm{Q}^{\mu \nu}_{\epsilon}(\Phi) ) &= \frac{1}{8 \pi} \bigg[h^{\sigma [\mu} \nabla_{\sigma} \xi^{\nu]}- \xi_{\sigma} \nabla^{[\mu} h^{\nu] \sigma}-\frac{1}{2} h \nabla^{[ \mu} \xi^{\nu ]}-\nabla^{[ \mu} \delta \xi^{\nu ]}  \nonumber \\
	&-( h F^{\mu \nu}+ 2\delta F^{\mu \nu}) (A_{\sigma}\xi^{\sigma}+\lambda)-2 F^{\mu \nu} \delta A_{\sigma}\xi^{\sigma}-2 F^{\mu \nu}(A_{\sigma} \delta\xi^{\sigma}+\delta\lambda) \bigg] .
\end{align}
Thus, by using Eq.(\ref{k expression}) we obtain the surface charge density
\begin{equation}\label{surface charge expression}
	\bm{k}_{\epsilon}(\delta \Phi,\Phi)=k^{\mu \nu}_{\epsilon}(\delta \Phi,\Phi) \sqrt{-g}(d^{n-2}x)_{\mu \nu} \, ,
\end{equation}
where
\begin{equation}\label{key}
	k^{\mu \nu}_{\epsilon}(\delta\Phi,\Phi)=k^{\mu \nu}_{g}(\delta\Phi,\Phi)+k^{\mu \nu}_{A}(\delta\Phi,\Phi),
\end{equation}
with
\begin{align}\label{scharge1}
	k^{\mu \nu}_{g}(\delta\Phi,\Phi) &= \frac{1}{8 \pi} \left[h^{\sigma [\mu} \nabla_{\sigma} \xi^{\nu]}- \xi_{\sigma} \nabla^{[\mu} h^{\nu] \sigma}-\frac{1}{2} h \nabla^{[ \mu} \xi^{\nu ]}+\xi^{[\mu} \nabla_{\sigma} h^{\nu] \sigma}-\xi^{[\mu} \nabla^{\nu]} h \right] , \nonumber \\
	k^{\mu \nu}_{A}(\delta\Phi,\Phi) &=\frac{-1}{4 \pi} \bigg[(\frac{1}{2} h F^{\mu \nu}+\delta F^{\mu \nu}) (A_{\sigma}\xi^{\sigma}+\lambda)+F^{\mu \nu} \delta A_{\sigma}\xi^{\sigma}+2\xi^{[ \mu} F^{\nu ] \sigma} \delta A_{\sigma} \bigg] .
\end{align}

\subsubsection{Conserved charges and the first law for a torus-like black hole}
The charged torus-like black hole solution to the Einstein-Maxwell EOMs (\ref{EOMs}) is given by \cite{Huang:1995zb}
\begin{align}\label{key}
	\mathrm{d} s^2 &=-f(r) \mathrm{d}t^2+\frac{\mathrm{d}r^2}{f(r)}+r^2(\mathrm{d}\theta^2+\mathrm{d}\varphi^2) , \nonumber \\
	f(r) &=-\frac{\Lambda r^2}{3}-\frac{2 m}{\pi r}+\frac{4 q^2}{\pi r^2} , \nonumber \\
	A_{\mu} &=(-\frac{2 q}{\sqrt{\pi} r}, ~0, ~0, ~0) ,
\end{align}
where $m$ and $q$ are the mass and electric charge parameters, respectively. The coordinates satisfy $0 \leqslant \theta \leqslant 2 \pi$ and $0 \leqslant \varphi \leqslant 2 \pi$. For negative $\Lambda$, this solution has two horizons: inner horizon $r_{-}$ and out horizon $r_{+}$, which solve $f(r_{\pm})=0$.

Since the integration in Eq.(\ref{eqdh}) is independent of the choice of the integration surface $\partial \Sigma$, we take $\partial \Sigma$ to be the torus of constant $(t, r)$ for simplicity and take the limit $r \rightarrow \infty$. Then, the conserved charge variation conjugate to the symmetry generator $\eta$ can be written as
\begin{equation}\label{deltaH}
	\delta H_{\eta}(\Phi)=\oint_{\partial \Sigma} \bm{k}_{\eta}(\delta \Phi, \Phi)=\int_{0}^{2 \pi} \int_{0}^{2 \pi} \lim_{r\to \infty}\sqrt{-g} \, k_{\eta}^{t r}(\delta \Phi, \Phi) \mathrm{d}\theta {} \mathrm{d}\varphi,
\end{equation}
where $k_{\eta}^{t r}$ is the $t r$ component of $k_{\eta}^{\mu \nu}$. The dynamical fields are $\Phi=(g_{\mu \nu}, A_{\mu})$, parametrized by $p_{\alpha}=\{m, q \}$. The parametric variations \cite{Hajian:2014twa} are given by
\begin{align}\label{parametric variations}
	\delta g_{\mu \nu} &=\frac{\partial g_{\mu \nu}}{\partial m} \delta m+\frac{\partial g_{\mu \nu}}{\partial q} \delta q ,\nonumber \\
	\delta A_{\mu} &=\frac{\partial A_{\mu}}{\partial m} \delta m+\frac{\partial A_{\mu}}{\partial q} \delta q .
\end{align}
\textbf{Mass:} By choosing the symmetry generator $\eta_{{}_M}=\{ \partial_{t}, 0 \}$ and substituting the surface charge density expressions (\ref{scharge1}) along with the parametric variations (\ref{parametric variations}) into Eq.(\ref{deltaH}), we can obtain the total mass
\begin{equation}\label{key}
	\delta M=\delta H_{\eta_{{}_M}}= \delta m  \quad \Rightarrow \quad M=m .
\end{equation}
\textbf{Electric charge:} By choosing the symmetry generator $\eta_{{}_Q}=\{ 0, -1 \}$, and using Eq.(\ref{deltaH}), the electric charge is given by
\begin{equation}\label{key}
	\delta Q=\delta H_{\eta_{{}_Q}}=2 \sqrt{\pi} \delta q    \quad \Rightarrow \quad Q=2 \sqrt{\pi} q .
\end{equation}
\textbf{Entropy:} The Killing vector generating the Killing horizon of the black hole is $\zeta = \partial_{t}$. The surface gravity and temperature on the horizon are defined by  \cite{Compere:2018aar}
\begin{align}
	\kappa_{{}_{\mathrm{H}}}&=\sqrt{-\frac{1}{2}(\nabla_{\mu}\zeta_{\nu})(\nabla^{\mu}\zeta^{\nu})}|_{r_{{}_{\mathrm{H}}}}=\frac{-12 q^2+3 m r_{{}_{\mathrm{H}}}-\pi \Lambda r_{{}_{\mathrm{H}}}^4 }{3 \pi r_{{}_{\mathrm{H}}}^3},  \nonumber\\
	T_{\mathrm{H}}&= \frac{\kappa_{{}_{\mathrm{H}}}}{2 \pi},  \nonumber\\
	\Phi_{\mathrm{H}}&=-A_{\mu}\zeta^{\mu}|_{r_{{}_{\mathrm{H}}}}=\frac{2  q}{\sqrt{\pi } r_{{}_{\mathrm{H}}}} ,
\end{align}
where $r_{{}_{\mathrm{H}}} \equiv r_{\pm}$. By choosing the symmetry generator $\eta_{{}_{\mathrm{H}}} = \frac{2 \pi}{\kappa_{{}_{\mathrm{H}}}} \{ \partial_{t}, \Phi_{\mathrm{H}} \}$ and using Eq.(\ref{deltaH}), the corresponding entropy variation is given by
\begin{equation}\label{key}
	\delta S=\delta H_{\eta_{{}_{\mathrm{H}}}}=\frac{2 \pi}{\kappa_{{}_{\mathrm{H}}}}(\delta m - \Phi_{\mathrm{H}} \cdot 2 \sqrt{\pi} \delta q) ,
\end{equation}
By the relations
\begin{equation}\label{key}
	\frac{\partial r_{{}_{\mathrm{H}}}}{\partial m}=\frac{6 r_{{}_{\mathrm{H}}}^2}{-24 q^2-2 \pi \Lambda r_{{}_{\mathrm{H}}}^4+6 m r_{{}_{\mathrm{H}}}} , \quad \frac{\partial r_{{}_{\mathrm{H}}}}{\partial q}=\frac{-24 q r_{{}_{\mathrm{H}}}}{-24 q^2-2 \pi \Lambda r_{{}_{\mathrm{H}}}^4+6 m r_{{}_{\mathrm{H}}}} ,
\end{equation}
we get
\begin{equation}\label{entropy}
	\delta S=2 \pi^2 r_{{}_{\mathrm{H}}} (\frac{\partial r_{{}_{\mathrm{H}}}}{\partial m}\delta m+\frac{\partial r_{{}_{\mathrm{H}}}}{\partial q}\delta q)=\delta(\pi^2 r_{{}_{\mathrm{H}}}^2) \quad \Rightarrow \quad S=\pi^2 r_{{}_{\mathrm{H}}}^2=\frac{\mathcal{A}}{4} ,
\end{equation}
where $\mathcal{A}=4 \pi^2 r_{{}_{\mathrm{H}}}^2$ is the horizon area of the torus-like black hole. \\
\textbf{The first law:} From the proof of Eq.(\ref{proof}), the first law is satisfied
\begin{equation}\label{The first law1}
	\delta S=\frac{1}{T_{\mathrm{H}}}(\delta M-\Phi_{\mathrm{H}} \delta Q).
\end{equation}     
The reference points for the charges above are chosen to vanish when $m=q=0$. 

In our results, the electric charge $Q$ differs from that obtained in the original literature \cite{Huang:1995zb}, where $Q=q$. For the torus-like black hole, the entropy is exactly one quarter of the horizon area and satisfies the first law. 

%%%%%%%%%%%%%%%%%%%%%%%%%%%%%%%%%%%%%%%%%%%%%%%%%%%%%%%%%%%%%%%%%%%%%%%%%%%%%%%%%%%%%%%%%%%%%%%%  
\subsection{Einstein-Euler-Heisenberg nonlinear electrodynamics}
In this subsection we consider the Einstein-Euler-Heisenberg nonlinear electrodynamics (EEH NED). The dynamics follow from the Lagrangian \cite{Magos:2020ykt}
\begin{equation}\label{LEEH}
	L=\frac{1}{16 \pi}(R-2 \Lambda -4 L(X,Y)),
\end{equation}
with
\begin{equation}\label{LNED}
	L(X,Y)=-X+\frac{\mathscr{A}}{2} X^2+\frac{\mathscr{B}}{2} Y^2 ,
\end{equation}
where $R$ is the Ricci scalar, $\Lambda$ is the cosmological constant, and $L(X, Y)$ is the NED Lagrangian which depends on the electromagnetic invariants, $X=\frac{1}{4}F_{\mu \nu} F^{\mu \nu},\,\,\,\, Y=\frac{1}{4}F_{\mu \nu} \star F^{\mu \nu}.\,\,\,\,\star F^{\mu \nu}=\frac{1}{2}\epsilon^{\mu \nu \rho \sigma}F_{\rho \sigma}$ denotes the dual of the electromagnetic field strength $ F_{\mu \nu}=\nabla_{\mu} A_{\nu}-\nabla_{\nu} A_{\mu}$, in which $ A_{\nu}$ is the electromagnetic vector potential, and $\epsilon_{\mu \nu \rho \sigma} =\sqrt{-g} \varepsilon_{\mu \nu \rho \sigma}$ denotes the Levi-Civita tensor. The Euler-Heisenberg parameters are $\mathscr{A}=\frac{8 \alpha^2}{45 m^4},\,\,\,\,\mathscr{B}=\frac{7 \alpha^2} {180 m^4}=\frac{7}{4} \mathscr{A}$, where $m$ is the electron mass, and $\alpha$ is the fine structure constant.

The variations of the Lagrangian (\ref{LEEH}) give the EOMs
\begin{align}\label{NEOMs}
	\mathrm{E}^{\mu \nu}_{g} &=\frac{1}{16 \pi} \left[-(G^{\mu \nu}+\Lambda g^{\mu \nu})
	-4(\frac{1}{2} L g^{\mu \nu}-\frac{1}{2} L_X F^{\mu \sigma}F^{\nu}{}_{\sigma}-\frac{1}{2} L_Y Y g^{\mu \nu} ) \right], \nonumber \\
	\mathrm{E}^{\mu}_{A} &=\frac{1}{16 \pi} \nabla_{\nu} \left[ -4 \mathcal{F}^{\mu \nu} \right] ,
\end{align}
where
\begin{align}\label{key}
	\mathcal{F}^{\mu \nu} &=L_X F^{\mu \nu}+ L_Y \star F^{\mu \nu} ,   \nonumber \\
	L_X &=\frac{\partial L(X,Y)}{\partial X}=-1+\mathscr{A} X ,   \nonumber \\
	L_Y &=\frac{\partial L(X,Y)}{\partial Y}=\mathscr{B} Y .
\end{align}

As in the previous section, the symplectic potential and the Noether-Wald charge density are given by
\begin{align}\label{key}
	&\Theta^{\mu}(\delta \Phi, \Phi) =\frac{1}{16 \pi} \left[ 2 \nabla^{[\nu } {h^{\mu ]}}_{\nu}-4 \mathcal{F}^{\mu \nu} \delta A_{\nu} \right] , \nonumber \\
	&\mathrm{Q}^{\mu \nu}_{\epsilon}(\Phi) =\frac{1}{16 \pi}\left[-2 \nabla^{[\mu} \xi^{\nu ] } -4 \mathcal{F}^{\mu \nu} (A_{\sigma}\xi^{\sigma}+\lambda) \right] .
\end{align}
Considering $\delta \epsilon \neq 0$, the variation of the Noether-Wald charge density is given by
\begin{align}\label{key}
	\delta( \sqrt{-g} \mathrm{Q}^{\mu \nu}_{\epsilon}(\Phi) ) &= \frac{1}{8 \pi} \bigg[h^{\sigma [\mu} \nabla_{\sigma} \xi^{\nu]}- \xi_{\sigma} \nabla^{[\mu} h^{\nu] \sigma}-\frac{1}{2} h \nabla^{[ \mu} \xi^{\nu ]}-\nabla^{[ \mu} \delta \xi^{\nu ]}  \nonumber \\
	&-( h \mathcal{F}^{\mu \nu}+ 2\delta \mathcal{F}^{\mu \nu}) (A_{\sigma}\xi^{\sigma}+\lambda)-2 \mathcal{F}^{\mu \nu} \delta A_{\sigma}\xi^{\sigma}-2 \mathcal{F}^{\mu \nu}(A_{\sigma} \delta\xi^{\sigma}+\delta\lambda) \bigg] .
\end{align}
By using Eq.(\ref{k expression}) we obtain the surface charge density
\begin{align}\label{key}
	k^{\mu \nu}_{\epsilon}(\delta\Phi,\Phi) &= \frac{1}{8 \pi} \bigg[h^{\sigma [\mu} \nabla_{\sigma} \xi^{\nu]}- \xi_{\sigma} \nabla^{[\mu} h^{\nu] \sigma}-\frac{1}{2} h \nabla^{[ \mu} \xi^{\nu ]}+\xi^{[\mu} \nabla_{\sigma} h^{\nu] \sigma}-\xi^{[\mu} \nabla^{\nu]} h  \nonumber \\
	&-( h \mathcal{F}^{\mu \nu}+ 2\delta \mathcal{F}^{\mu \nu}) (A_{\sigma}\xi^{\sigma}+\lambda)-2 \mathcal{F}^{\mu \nu} \delta A_{\sigma}\xi^{\sigma}-4\xi^{[ \mu} \mathcal{F}^{\nu ] \sigma} \delta A_{\sigma} \bigg] .
\end{align}

\subsubsection{Conserved charges and the first law for an Einstein-Euler-Heisenberg AdS black hole}
To check the validity of our formalism, we study the conserved charges and the first law of thermodynamics for a charged EEH-AdS black hole in EEH NED. The charged EEH-AdS black hole solution to the EOMs (\ref{NEOMs}) is given by \cite{Magos:2020ykt} 
\begin{align}\label{key}
	\mathrm{d} s^2 &=-f(r) \mathrm{d}t^2+\frac{\mathrm{d}r^2}{f(r)}+r^2(\mathrm{d}\theta^2+\sin^2\theta \,\mathrm{d}\varphi^2) , \nonumber \\
	f(r) &=1-\frac{2 m}{r}+\frac{q^2}{r^2}-\frac{\Lambda r^2}{3}-\frac{\mathscr{A} q^4}{20 r^6} , \nonumber \\
	A_{\mu} &=(\frac{q}{ r}-\frac{\mathscr{A} q^3}{10 r^5}, ~0, ~0, ~0) ,
\end{align}
where $m$ and $q$ are the mass and electric charge parameters, respectively. The coordinates satisfy $0 \leqslant \theta \leqslant \pi$ and $0 \leqslant \varphi \leqslant 2 \pi$. The event horizon $r_{+}$ is determined by $f(r_{+})=0$. 

As in the previous subsection, for the charge variation in Eq.(\ref{eqdh}), we take $\partial \Sigma$ to be the sphere of constant $(t, r)$ and take the limit $r \rightarrow \infty$. Then, the conserved charge variation conjugate to the symmetry generator $\eta$ takes the form
\begin{equation}\label{NdeltaH}
	\delta H_{\eta}(\Phi)=\oint_{\partial \Sigma} \bm{k}_{\eta}(\delta \Phi, \Phi)=\int_{0}^{2 \pi} \int_{0}^{\pi} \lim_{r\to \infty}\sqrt{-g} \, k_{\eta}^{t r}(\delta \Phi, \Phi) \mathrm{d}\theta {} \mathrm{d}\varphi,
\end{equation}    
where the parametric variations $\delta \Phi$ for the dynamical fields $\Phi=(g_{\mu \nu}, A_{\mu})$ take the same form as Eq.(\ref{parametric variations}). \\
\textbf{Mass:} By choosing the symmetry generator $\eta_{{}_M}=\{ \partial_{t}, 0 \}$ and using Eq.(\ref{NdeltaH}), we can obtain the total mass
\begin{equation}\label{key}
	\delta M=\delta H_{\eta_{{}_M}}= \delta m  \quad \Rightarrow \quad M=m .
\end{equation}
\textbf{Electric charge:} Choosing the symmetry generator $\eta_{{}_Q}=\{ 0, -1 \}$ gives
\begin{equation}\label{key}
	\delta Q=\delta H_{\eta_{{}_Q}}=\delta q   \quad \Rightarrow \quad Q=q .
\end{equation}
\textbf{Entropy:} The Killing vector generating the Killing horizon of the black hole is $\zeta = \partial_{t}$. The surface gravity, temperature and electric potential on the horizon are defined by \cite{Compere:2018aar}
\begin{align}
	\kappa_{{}_{\mathrm{H}}}&=\sqrt{-\frac{1}{2}(\nabla_{\mu}\zeta_{\nu})(\nabla^{\mu}\zeta^{\nu})}|_{r_{{}_{\mathrm{H}}}}=\frac{1}{2r_{{}_{\mathrm{H}}}}(1-\frac{q^2}{r_{{}_{\mathrm{H}}}^2}-\Lambda r_{{}_{\mathrm{H}}}^2+\frac{\mathscr{A} q^4}{4 r_{{}_{\mathrm{H}}}^6} ),  \nonumber\\
	T_{\mathrm{H}}&= \frac{\kappa_{{}_{\mathrm{H}}}}{2 \pi},  \nonumber\\
	\Phi_{\mathrm{H}}&=-A_{\mu}\zeta^{\mu}|_{r_{{}_{\mathrm{H}}}}=-\frac{q}{r_{{}_{\mathrm{H}}}}+\frac{\mathscr{A} q^3}{10 r_{{}_{\mathrm{H}}}^5},
\end{align}
where $r_{{}_{\mathrm{H}}} \equiv r_{+}$. By choosing the symmetry generator $\eta_{{}_{\mathrm{H}}} = \frac{2 \pi}{\kappa_{{}_{\mathrm{H}}}} \{ \partial_{t}, -\Phi_{\mathrm{H}} \}$, we obtain the entropy
\begin{equation}\label{Nentropy}
	\delta S=\delta H_{\eta_{{}_{\mathrm{H}}}}=\delta(\pi r_{{}_{\mathrm{H}}}^2) \quad \Rightarrow \quad S=\pi r_{{}_{\mathrm{H}}}^2 .
\end{equation} 
\textbf{The first law:} From the proof of Eq.(\ref{proof}), the first law is satisfied
\begin{equation}\label{NThe first law1}
	\delta S=\frac{1}{T_{\mathrm{H}}}(\delta M-\Phi_{\mathrm{H}} \delta Q).
\end{equation}
Still, we choose the reference points for the charges above to vanish when $m=q=0$. 

Although the NED in the CPSF was studied in Ref. \cite{Bokulic:2021dtz}, our formulation includes the internal gauge transformation, and we provide an explicit example to illustrate its validity. From the investigation above, we can see that the mass and the electric charge are not affected compared with the original Wald’s formalism, since the symmetry generators of mass and electric charge are field independent, i.e., $\delta \eta_{{}_M} = 0$ and $\delta \eta_{{}_Q} = 0$. The field dependence of conserved charges just presented in the entropy by $\delta \eta_{{}_{\mathrm{H}}} \neq 0$. By considering the field dependence of the symmetry generators, the first law of black hole thermodynamics is a direct result for generic black hole solutions in generally invariant gravitational theories.

\section{Conclusions and Outlook}
In this paper, we generalize the CPSF conjugate to field dependent vectors \cite{Compere:2015knw} to include internal gauge transformations. The symmetry generator is a combination of diffeomorphism plus internal gauge transformation $\epsilon = \left\{ \xi, \lambda \right\}$, and the general gauge transformation takes the form $\delta_{\epsilon} \Phi = \left\{ \mathscr{L}_{\xi} \Phi , \delta_{\lambda} \Phi \right\}$. More generally, we consider field dependent symmetry generators $\epsilon=\epsilon (\Phi)$, which implies $\delta \epsilon \neq 0$. When the field dependence of the symmetry generators is considered in the CPSF, the first law of black hole thermodynamics is a direct result for generally invariant gravitational theories (up to total derivative terms). Since the symmetry generators of mass, angular momentum and electric charges are field independent, these conserved charges are unchanged from those obtained using the original Wald’s formalism, which doesn’t account for the field dependence of symmetry generators. The field dependence of the conserved charges is only presented in entropy. In our formalism, the integration surfaces are arbitrary and independent of any chosen horizon or asymptotics. Only the chemical potentials $ \left(T_{\mathrm{H}}, \Omega^i_{\mathrm{H}}, \Phi^a_{\mathrm{H}}, \cdots \right) $ are evaluated at the horizons. To check the validity of our formalism, we investigate the conserved charges and the first laws of thermodynamics for the torus-like black hole in Einstein-Maxwell theory and the charged EEH-AdS black hole in EEH NED. Our formalism is universal and can be applied to any generic black hole solution in generally invariant gravitational theories (up to total derivative terms).

It is straightforward to generalize our formalism to non-Abelian Einstein-Yang-Mills theory with the gauge transformation $\delta_{\lambda} \mathbf{A}=\mathrm{d} \bm{\lambda}+\left[ \mathbf{A}, \bm{\lambda} \right]$, where $\mathbf{A}$ is the Yang-Mills potential \cite{Klinger:2023qna}, as well as to include an $(n-1)$-form gauge field $\mathbf{A}^a$ with the gauge transformation $\delta_{\bm{\lambda}^{a}} \mathbf{A}^{a}=\mathrm{d} \bm{\lambda}^{a} $ for an $(n-2)$-form gauge symmetry generator $\bm{\lambda}^{a}$ \cite{Chernyavsky:2017xwm,Hajian:2023bhq}.

%######################################################################################%

%######################################################################################%

~\\
\noindent \textbf{Acknowledgements}\\
The author would like to thank Dr. Feiyi Liu for helpful discussions. This work is supported partially by Yunnan Provincial Department of Education Science Research Fund Project (Grant No. 2025J0946).

\end{document}